# Deep learning for bioimage analysis

Adrien Hallou[=,1,2,3], Hannah Yevick[=,4], Bianca Dumitrascu[✉,5] & Virginie Uhlmann[✉,6]

1. Cavendish Laboratory, Department of Physics, University of Cambridge, Cambridge, CB3 0HE, UK.
2. Wellcome Trust/Cancer Research UK Gurdon Institute, University of Cambridge, Cambridge, CB2 1QN, UK.
3. Wellcome Trust/Medical Research Council Stem Cell Institute, University of Cambridge, Cambridge, CB2 1QR, UK.
4. Department of Biology, Massachusetts Institute of Technology, Cambridge, MA, 02142, USA.
5. Computer Laboratory, Cambridge, University of Cambridge, Cambridge, CB3 0FD, UK.
6. European Bioinformatics Institute, European Molecular Biology Laboratory, Cambridge, CB10 1SD, UK.

## Key words

Deep learning, neural network, image analysis, microscopy, bioimaging.

## Summary statement

This Review summarizes recent advances in bioimage analysis enabled by deep learning algorithms.

## Abstract

Deep learning has transformed the way large and complex image datasets can be processed, reshaping what is possible in bioimage analysis. As the complexity and size of bioimage data continues to grow, this new analysis paradigm is becoming increasingly ubiquitous. In this Review, we begin by introducing the concepts needed for beginners to understand deep learning. We then review how deep learning has impacted bioimage analysis and explore the open-source resources available to integrate it into a research project. Finally, we discuss the future of deep learning applied to cell and developmental biology. We analyse how state-of-the-art methodologies have the potential to transform our understanding of biological systems through new image-based analysis and modelling that integrate multimodal inputs in space and time.

## Introduction

In the past decade, deep learning (DL) has revolutionized biology and medicine through its ability to automate repetitive tasks and integrate complex collections of data to produce

---

[=] A.H and H.Y contributed equally to this work.
[✉] To whom correspondence may be addressed. Email: bmd39@cam.ac.uk & uhlmann@ebi.ac.uk .

reliable predictions (LeCun et al., 2015). Among its many uses, DL has been fruitfully exploited for image analysis. While the first DL approaches successfully used for the analysis of medical and biological data were initially developed for computer vision applications such as image database labelling (Krizhevsky et al., 2012), many research efforts have since then focused on tailoring DL for medical and biological image analysis (Litjens et al., 2017), understood as the computer-based analysis of microscopy images of biological objects. Bioimages (see Glossary, Box 1), in particular, exhibit a large variability due to the countless different possible combinations of phenotypes of interest, sample preparation protocols, imaging modalities and acquisition parameters. DL is thus a particularly appealing strategy to design general algorithms that can easily adapt to specific microscopy data with minimal human input. For this reason, the successes and promises of DL in bioimage analysis applications have been the topic of a number of recent review articles (Gupta et al., 2018; Wang et al., 2019; Moen et al., 2019; Meijering, 2020; Hoffmann et al., 2021; Esteva et al., 2021).

Here, we expand upon a recent Spotlight article (Villoutreix, 2021) and tour the practicalities of the use of DL for image analysis in the context of developmental biology. We first provide a primer on key machine learning and DL concepts. We then review the use of DL in bioimage analysis and outline success stories of DL-enabled bioimage analysis in developmental biology experiments. For readers wanting to further experiment with DL, we compile a list of freely available resources, most requiring little to no coding experience. Finally, we discuss more advanced DL strategies that are still under active investigation but are likely to become routinely used in the future.

## What is machine learning?

The term machine learning (ML) defines a broad class of statistical models and algorithms that allow computers to perform specific data analysis tasks. Examples of tasks include, but are not limited to, classification, regression, ranking, clustering or dimensionality reduction [defined in (Mohri et al., 2018)], and are usually performed on datasets collected with or without prior human annotations.

Three main ML paradigms can be distinguished: supervised, unsupervised and reinforcement learning (Murphy, 2012; Villoutreix, 2021). The overwhelming majority of established bioimage analysis algorithms rely on the two first ones, and we therefore focus on supervised and unsupervised ML in the rest of the article. In supervised learning, prior human knowledge is used to obtain a 'ground truth' label for each element in a dataset. The resulting data-label pairs are then split into a 'training' and a 'testing' set (see Glossary, Box 1). Using the training set, the ML algorithm is 'trained' to learn the relationship between 'input' data and 'output' labels by minimizing a 'loss' function (see Glossary, Box 1), and its performance is assessed on the testing set. Once training is complete, the ML model can be applied to unseen, but related, input data in order to predict output labels. Classical supervised ML methods include random forests, gradient boosting, and support vector machines (Mohri et al., 2018). In contrast, unsupervised learning deals with unlabelled data: ML is then employed to uncover patterns in input data without human-provided examples. Examples of unsupervised learning tasks include clustering and dimensionality reduction, as routinely used for instance in the analysis of single cell omics data (Libbrecht & Noble 2015; Argelaguet et al., 2021).

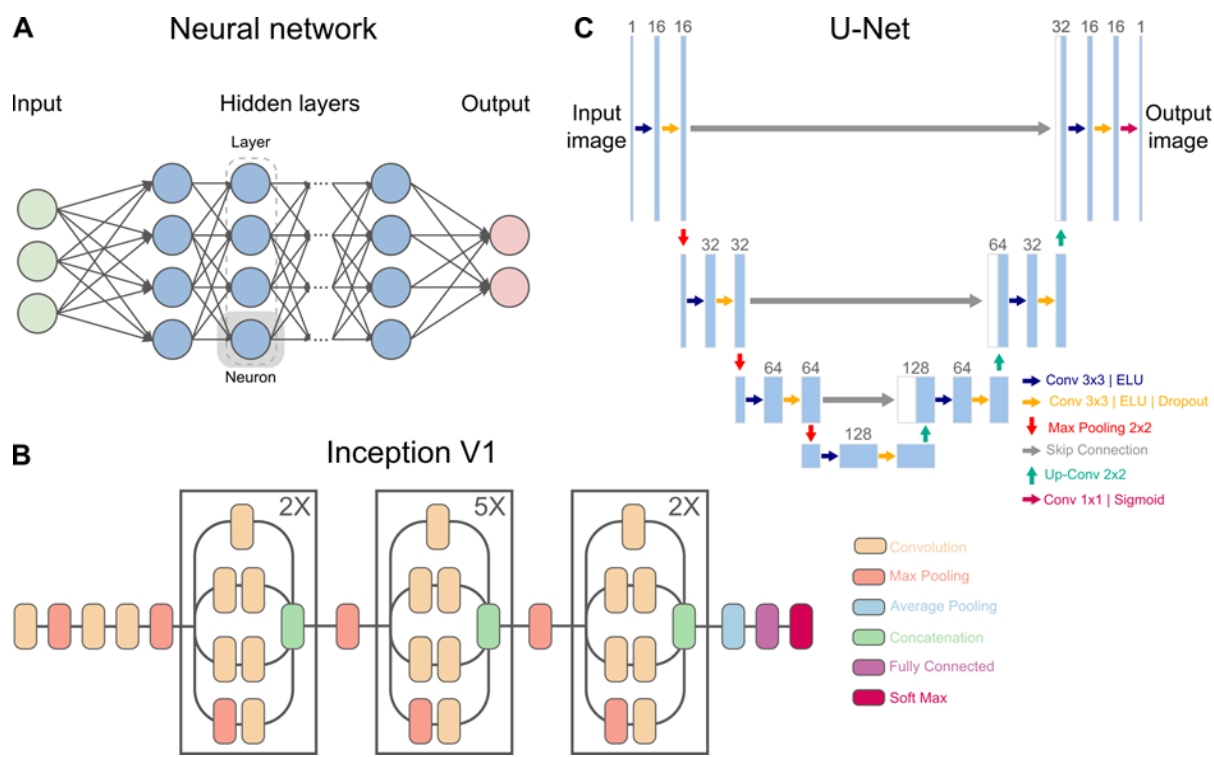

**Fig. 1. Neural networks and convolutional neural networks for bioimage analysis. (A)** Schematic of a typical NN composed of an input layer (green), hidden layers (blue) and an output layer (red). Each layer is composed of neurons connected to each other. **(B)** Schematic of an Inception V1 architecture, also called "GoogleLeNet". Inception V1 is a typical CNN architecture for image classification tasks. It has for example been used to classify early human embryos images with very high accuracy (Khosravi et al., 2019). It is designed around a repetitive architecture made of so-called "inception blocks", which apply several 'convolutional' and max 'pooling' layers to their input prior to concatenate together all generated feature maps (see Glossary, Box 1). **(C)** Schematic of a U-net architecture as used in (McGinn et al., 2021) for segmentation of cells and nuclei in mouse epithelial tissues. U-net is amongst the most popular and efficient CNN models used for bioimage analysis and is designed using 'convolutional', 'pooling' and 'dense' layers as key building blocks (see Glossary, Box 1). U-net follows a symmetric encoder-decoder architecture resulting in a characteristic U-shape. Along the encoder path, the first branch of the U, the input image is progressively compacted, leading to a representation with reduced spatial information but increased feature information. Along the decoder path, the second branch of the U, the feature and spatial information are combined with information from the encoder path, enforcing the model to learn image characteristics at various spatial scales.

## Neural networks and deep learning

DL designates a family of ML models based on neural networks (NN) (LeCun et al. 2015). Formally, a NN aims to learn non-linear maps between inputs and outputs. A NN is a network of processing 'layers' composed of simple, but non-linear, units called artificial neurons (see Glossary, Box 1). When composed of several layers, a NN is referred to as a deep NN. Layers of artificial neurons transform inputs at one level (starting with input data) into outputs at the next level such that the data becomes increasingly more abstract as it progresses through the different layers, encapsulating in the process the complex non-linear relationship usually existing between input and output data (Fig. 1A). This process allows sufficiently deep NN to learn during training some higher-level features contained in the data (Goodfellow et al., 2016). For example, for a classification problem such as the identification of cells contained in a fluorescence microscopy image, this would typically involve learning features correlated with cell contours, while ignoring the noisy variation of pixel intensity in the background of the image.

# Box 1. Glossary of technical terms

**Bioimage:** visual observations of biological structures and processes at various spatiotemporal resolutions stored as digital image data.

**Ground truth:** output known to be correct for a given input.

**Training/testing set:** collections of known input-output pairs. The training set is used during training per se, while the testing set is used a posteriori to test the performances of the ML model on unseen data.

**Training**: process through which an ML model's parameters are optimized to best map inputs into desired outputs.

**Input data**: data fed into an ML model.

**Output data**: data coming out of an ML model.

**Loss:** function evaluating how closely the predictions of a model match the ground truth.

**Layer:** set of interconnected artificial neurons in a NN.

**Weights:** NN parameters that are iteratively adjusted during the training process.

**Accuracy:** ratio of correctly predicted instances to the total number of predicted instances.

**Signal-to-noise ratio** or **SNR:** measure of image quality usually computed as the ratio of the mean intensity value of a digital image to the standard deviation of its intensity values.

**Image patch:** small, rectangular piece of a larger image (e.g., 64x64 pixels patches for 1024x1024 pixels images) used to minimize computational costs during training.

**U-net:** a highly efficient CNN architecture used for various image analysis tasks (Fig. 1C).

**Transfer learning:** method in which a ML model developed for a task is reused for a different task. For instance, an NN can be initialized with the weights of another NN pre-trained on a large unspecific image dataset, and then fine-tuned with a problem-specific training set of smaller size.

**Style transfer:** method consisting of learning a specific style from a reference image, such that any input image can then be 'painted' in the style of the reference while retaining its specific features.

**Data augmentation:** strategy to enhance the size and quality of training sets. Typical techniques include random cropping, geometrical operations (e.g., rotations, translations, flips), intensity and contrast modifications, and non-rigid image transformations (e.g., elastic deformations).

> **Convolutional layer:** a type of layer akin to an image processing filter whose values are free parameters to be learnt during training. Each neuron in a convolutional layer is only connected to a few adjacent neurons in the previous layer.
>
> **Pooling:** operation consisting of aggregating adjacent neurons with a maximum, minimum, or averaging operator.
>
> **Dense layer** or **fully connected layer**: a type of layer in which all neurons are connected to all the neurons in the preceding layer.

Intuitively, a DL model can be viewed as a machine with many tuneable knobs, which are connected to one another through links. Tuning a knob changes the mathematical function that transforms the inputs into outputs. This transformation depends on the strength of the links between the knobs, and the importance of the knobs, known together as 'weights' (see Glossary, Box 1). A model with randomly set weights will make many mistakes, but the so-called 'winning lottery hypothesis' (Frankle & Carbin, 2018) assumes that an optimal configuration of knobs and weights exists. This optimal configuration is searched for during training, in which the knobs of the DL model are reconfigured by minimizing the loss function. While prediction with trained networks is generally fast, training deep NN *de novo* proves more challenging. A main difficulty in DL lies in finding an appropriate numerical scheme that allows, with limited computational power, tuning the tens of thousands of weights contained in each layer of the networks and obtaining high accuracy (see Glossary, Box 1) (LeCun et al., 2015; Goodfellow et al., 2016). While the idea (McCulloch & Pitts, 1943) and the first implementations of NN (Rosenblatt, 1958) date back to the dawn of digital computing, it took several decades for the development of computing infrastructure and efficient optimisation algorithms to allow implementations of practical interest, such as handwritten-digit recognition (LeCun et al., 1989).

Convolutional NN (CNN) are a particular type of NN architecture specifically designed to be trained on input data that are multidimensional arrays such as images. CNN attracted particular interest in image processing when, in the 2012 edition of the ImageNet challenge on image classification, the AlexNet model outperformed by a comfortable margin other ML algorithms (Krizhevsky et al., 2017). In bioimage analysis application, the U-net architecture (Falk et al., 2019) (Fig. 1C) has become predominant, as further discussed in the next section.

## Deep learning for bioimage analysis

DL in bioimage analysis tackles three main kind of tasks: (i) image restoration, where an input image is transformed into an enhanced output image; (ii) image partitioning, whereby an input image is divided into regions and/or objects of interest; and (iii) image quantification, whereby objects are classified, tracked, or counted. Here, we illustrate each class of applications with examples of DL-enabled advances in cell and developmental biology.

### Image restoration

Achieving a high signal-to-noise ratio (SNR; see Glossary, Box 1) when imaging an object of interest is a ubiquitous challenge when working with developmental systems. Noise in microscopy can arise from several sources (e.g. the optics of the microscope and/or its associated detectors or camera). Live imaging, in particular, usually compromises between SNR, acquisition speed, and imaging resolution. In addition, areas of interest in developing systems are frequently embedded inside the organism, far from the microscope objective. Therefore, due to scattering, light traveling from fluorescent markers can be distorted and of decreased intensity when it reaches the objective. Photobleaching and phototoxicity are also increasingly problematic deeper into the tissue, leading to low SNR as one mitigates its effect through decreased laser power and increased camera exposure or detector voltage. DL has been successful at overcoming these challenges when used in the context of image restoration algorithms, which transform input images into output images with improved SNR.

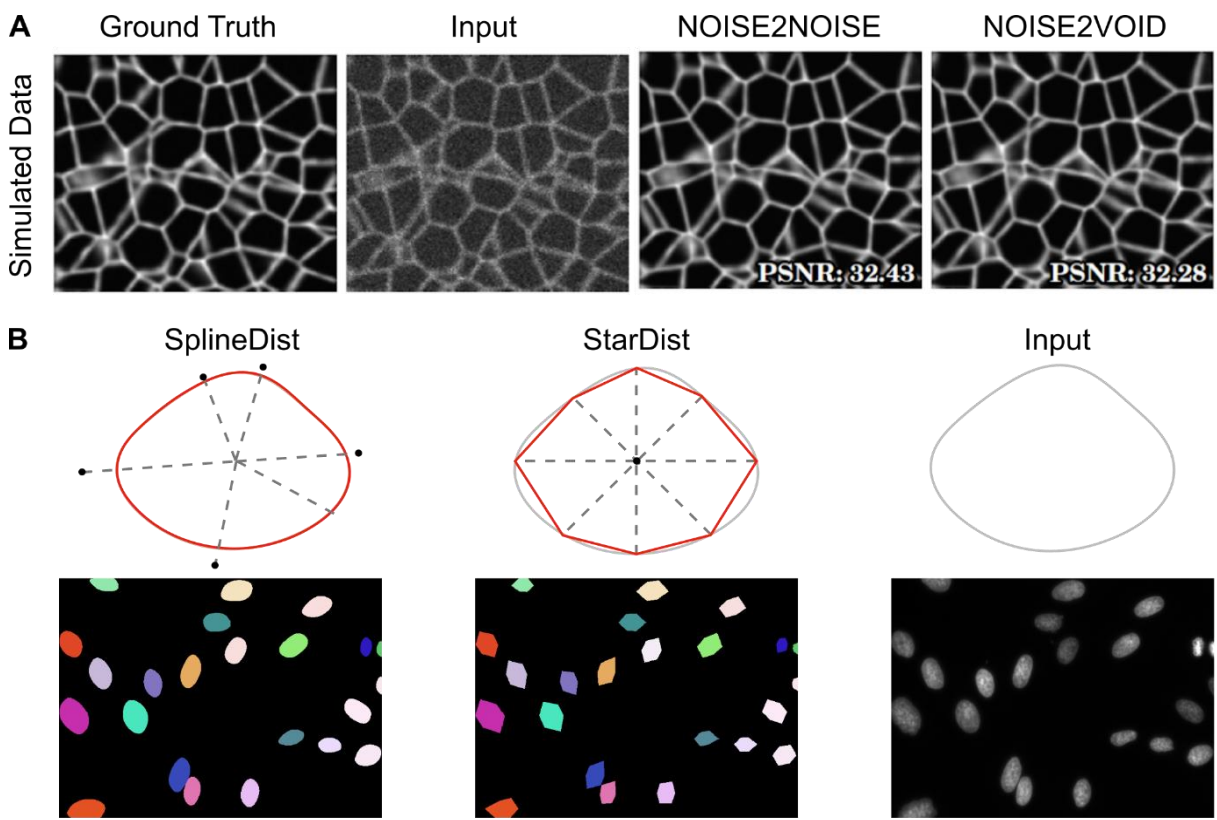

**Fig 2. Deep learning methods applied to developmental biology applications. (A)** A simulated ground truth cell membrane image is artificially degraded with noise. Denoised outputs obtained using Noise2Noise and Noise2Void are shown on the right, along with their average peak signal-to-noise values (higher values translate to sharper, less noisy images). Image adapted from (Krull et al., 2019). **(B)** Fluorescence microscopy cell nuclei image from the Kaggle 2018 Data Science Bowl, dataset: BBBC038v1 (Caicedo et al. 2019) segmented with StarDist (Schmidt et al., 2018), in which objects are represented as star-convex polygons, and with SplineDist, in which objects are described as a planar spline curve. Image adapted from (Mandal et al., 2021).

Although algorithms relying on theoretical knowledge of imaging systems have made image restoration possible since the early days of bioimage analysis (Born et al., 2003; Gibson & Lanni, 1989; Gibson & Lanni, 1992), the competitive performance of both supervised and unsupervised forms of DL has introduced a paradigm shift. Despite lacking in theoretical guarantees, several purely data-driven DL-based approaches outshine non-DL strategies in accurate image restoration tasks. One challenge in applying supervised DL to image restoration is the need for high-quality training sets of ground truth images exhibiting a reduced amount of noise. A notable example of DL-based image restoration algorithm requiring a relatively small training set [200 image patches (see Glossary, Box 1), size 64×64×16 pixels] is content-aware image restoration (CARE) (Weigert et al., 2018). To train CARE, pairs of registered low-SNR and high-SNR images must first be acquired. The high-SNR images serve as ground truth for training a DL model based on the U-net architecture (Falk et al., 2019) (see Glossary, Box 1; Fig. 1C). The trained network can then be used to restore noiseless, higher-resolution images from unseen noisier datasets (Box 2). Often, however, high-SNR ground truth image data cannot be easily generated experimentally. In such cases, synthetic high-SNR images generated by non-DL deconvolution algorithms can be used to train the network. For example, CARE has been trained to resolve sub-diffraction structures in low-SNR brightfield microscopy images using synthetically generated super-resolution data (Weigert et al., 2018). More recently, the DECODE method (Speiser et al., 2020) uses a U-net architecture to address the related challenge of computationally increasing resolution in the context of single molecule localization microscopy. The U-net model takes into account multiple image frames, as well as their temporal context. DECODE can localize single fluorophore emitters in 3D for a wide range of emitter brightnesses and densities, making it more versatile compared to previous CNN-based methods (Nehme et al. 2020, Boyd et al. 2018).

Unsupervised methods for image restoration offer an alternative to the generation of dedicated or synthetic training sets. Some recent denoising approaches exploit DL to learn how to best separate signal (e.g. the fluorescent reporter from a protein of interest) from noise, in some cases without the need for any ground truth. Noise2Noise, for instance, uses a U-net model to restore noiseless images after training on pairs of independent noisy images, and was demonstrated to accurately denoise biomedical image data (Lehtinen et al., 2018) (Fig. 2A). Going further, Noise2Self modifies Noise2Noise to only require noisy images split into input and target sets (Batson & Royer, 2019). In these algorithms, training is carried out on noisy images under the assumption that noise is statistically independent in image pairs, while the signal present is more structured. Alternatively, Noise2Void proposes a strategy to train directly on the dataset that needs to be denoised (Krull et al., 2019) (Fig. 2A). The Noise2 model family is ideal for biological applications in which it can be challenging to obtain noise-free images.

## Image partitioning

Analyzing specific objects in a biological image generally requires an image partitioning step, i.e., the separation of objects of interest from the image background. Image partitioning can either consist of detecting a bounding box around objects (object detection), or of identifying the set of pixels composing each object (segmentation). While images featuring a few objects can be partitioned by hand, large datasets necessitate automation. DL approaches originating from computer vision have greatly enhanced the speed and accuracy of both object detection and segmentation in biological images. Since U-net, countless customized DL models adapted to bioimage-specific object detection (Waithe et al., 2020; Wollman & Rohr, 2021) and segmentation problems have been proposed (Long, 2020; Chidester et al., 2019; Tokuoka et al., 2020). A strong link to computer vision remains, as many of these methods draw from partitioning tasks in natural images. For instance, algorithms initially designed to segment people and cars from crowded cityscapes can be efficiently exploited to segment challenging electron microscopy datasets (Wolf et al., 2018; Wolf et al., 2020). The automated segmentation of cell nuclei in various kinds of microscopy images has attracted a particular amount of attention. Cell nuclei can be tightly packed, making nuclei and cell bodies poorly differentiable from neighbours. Spatial variations in marker intensity due to local differences in staining efficacy, chromatin compaction or illumination fluctuations introduce further challenges. Mask Region-based CNN (He et al., 2017), an extension of Fast R-CNN, first developed for the general task of object detection in natural images, has been successfully adapted to nuclei segmentation. Building on this, StarDist (Schmidt et al., 2018) adds assumptions about the geometry of nuclei shapes to improve detection performance.

**Box 2. Case Study: Denoising the lateral cell faces of the developing *Drosophila* wing disc with CARE**

(Sui et al., 2018) and (Sui et al., 2020) explore the role of lateral tension in the *Drosophila* wing disc in guiding epithelial folding. The plane of the fly wing is mounted facing the objective, placing the lateral sides of the cell along the z-axis. However, image resolution in the (x-y) plane of a microscope (top left) generally exceeds that of out-of-plane (z) resolution (bottom left). Reconstructing fluorescent signals from the lateral face also requires reconstruction of z profiles by summing together signals from multiple depths. Furthermore, a sensitivity to light exposure of the system imposes that imaging be carried out at low laser power and on a few z slices, further decreasing the resolution of the lateral face. The quality of the acquired microscopy data is successfully improved relying on CARE (Weigert et al., 2018). The CARE network is trained on pairs of low and high resolution imaginal discs images, first a z stack is acquired using low laser power and low z sampling, followed by another z stack acquired at the same position in the sample with increased laser power and 4× more imaged focal planes (N=8 stacks of average dimensions 102×512×30 with pixel size 0.17×0.17×0.32 μm, for a total dataset size of ~1GB). Once trained, the network is used to process low-resolution images of other lateral markers, enabling the quantitative analysis of how protein localization changes over time on lateral cell faces during and after photoactivation. While absolute intensity measurements extracted from images restored with DL methods should be subject to caution, restored images in this work were only used to track relative changes in apical, basal and lateral intensity over time (bottom right). Scale bar 10.0 μm. Image adapted from (Sui et al. 2018).

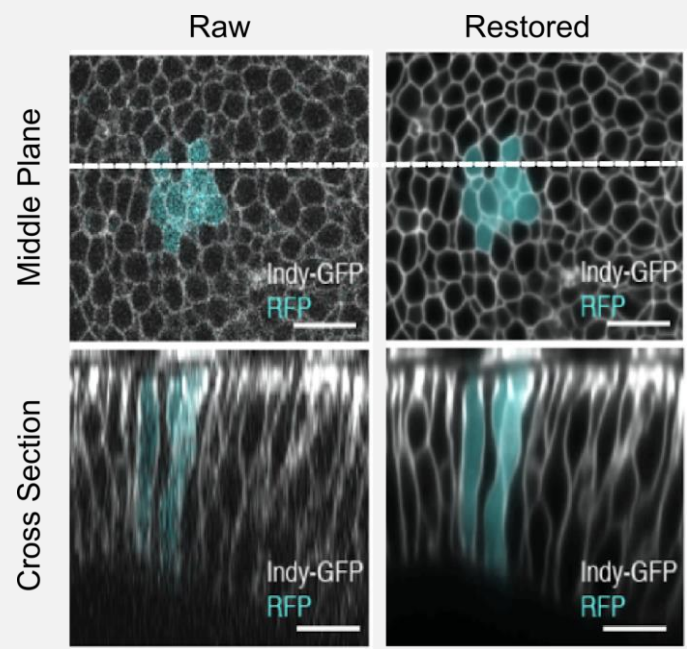

Relying on a U-net model, StarDist predicts a star-convex representation of individual object contours and can successfully separate overlapping nuclei in 2D images (Fig. 2B). A 3D version of StarDist is also available for volumetric (e.g. lightsheet microscopy) data as often encountered in developmental biology experiments (Weigert et al., 2020). More recently, SplineDist extends StarDist by using a more flexible representation of objects, allowing for the segmentation of more complex shapes (Mandal et al., 2021) (Fig. 2B). For these methods, larger training sets and crowdsourced improvements on model architecture have pushed the limits of achievable accuracy and generalization.

### Box 3. Case study: Automatic whole cell organelle segmentation in volumetric electron microscopy

Reconstructing the shape of internal components from focused ion beam scanning electron microscopy (FIB-SEM) data is a complicated task due to the crowded cytoplasmic environment of a cell. As a result, segmentation has been a bottleneck for understanding organelle morphologies and their spatial interactions as observed in SEM images at the nanometer scale. In OpenOrganelle (Heinrich et al., 2020), an ensemble of 3D U-nets has been trained for organelle segmentation in diverse cell types. The model is able to segment and classify up to 35 different classes of organelle, ranging from endoplasmic reticulum to microtubules to ribosomes. The network is trained with a diverse dataset of 73 volumetric regions. The image volumes are sampled from five different cell types, summing up to approximately $635 \times 10^6$ voxels. The identity of enclosed organelles in the chosen volumes are manually annotated using morphological features established in the literature (A). Achieving a manual segmentation of the dense array of organelles in a single ~1 µm$^2$ FIB-SEM slice required two weeks of manual labour for an expert, meaning that manual annotation of an entire cell (2250×larger) would take ~60 years. In contrast, the DL model trained on these manual annotations is able to segment individual organelles on a whole cell volume in a matter of hours (B). Image adapted from (Heinrich et al., 2020).

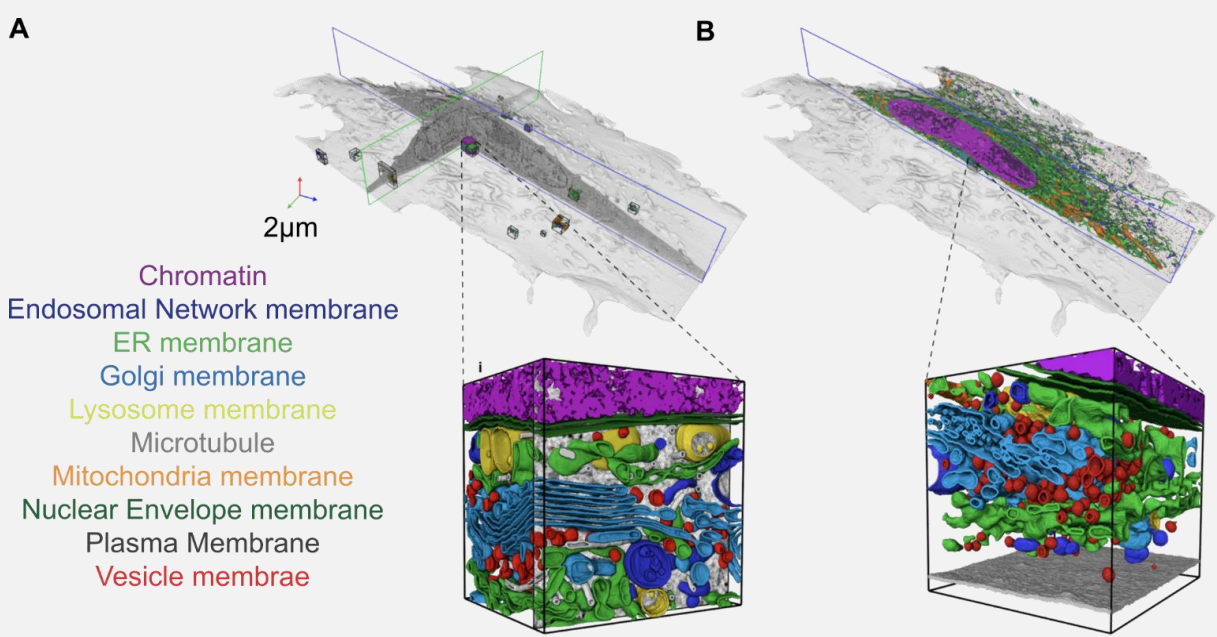

The availability of a benchmark dataset dedicated to nuclei segmentation has played a crucial role in this success. The 2018 Kaggle Data Science Bowl dataset (Caicedo et al., 2019), hosted as part of the Broad Bioimage Benchmark Collection, was assembled to faithfully reflect the variability of nuclei appearance and 2D image types in bioimaging. This large dataset was designed to challenge the generalization capabilities of segmentation methods across these variations and has established itself as a precious resource to objectively rank and comparatively assess algorithm performances. An equivalent 3D or 3D+time benchmark dataset is yet to be assembled and would be highly valuable to developmental biology image datasets, which are often volumetric and include a temporal component.

Cell membrane segmentation poses a more complex challenge than nuclei. Cells can take on varying morphologies, ranging from highly-stereotyped shapes to widely-varying sizes and contour roughness. DL models trained on a single dataset therefore often fail to infer accurately on different images. The true limits of the generalization capabilities of an algorithm is furthermore hard to assess in the absence of an established benchmarking dataset dedicated to whole cell segmentation. Cellpose (Stringer et al., 2021) takes on the generalization challenge relying on a large custom training set of microscopy images featuring cells with a wide range of diverse morphologies. This method relies on a U-net model predicting the directionality of spatial gradients in the input images and can process both 2D and 3D data. As a result, Cellpose is generalist enough to segment cells with very different morphologies and has been extensively reused (Young et al., 2021; Henninger et al., 2021). In addition, Cellpose is periodically re-trained with user-submitted data to continuously improve its performances (https://cellpose.readthedocs.io).

Finally, many types of biological questions require organelle segmentation. Manually segmenting organelles from 3D scanning electron microscope (SEM) images is highly time consuming, with annotating a single cell estimated to take ~60 years (Heinrich et al., 2020) (Box 3). Here, DL has been transformative as well, making it possible to automate the segmentation and classification of a wide range of cellular structures.

Image quantification

Once objects have been detected in individual images, the subsequent step is their quantification. Quantification can be about the number of objects (counting), their type (categorization), their shape (morphometry) or their dynamics (tracking), among many others.

Categorization can either be done holistically for an entire object (e.g. wild-type versus mutant), or by looking at a specific aspect of an object (e.g. the shape of internal components). Manual object categorization is both time consuming and has the potential for bias even when carried out by experts. In addition to speeding up the process, DL-powered image classification can limit annotation variability. Visually assessing embryo quality, for example, is subject to dispute between embryologists (Paternot et al 2019). (Khosravi et al., 2019) built a DL classifier of early human embryos quality trained on 'good quality' and 'poor quality' labeled embryos that corresponded to the score given by the majority vote of five embryologists. Their model, based on Google's Inception-V1 architecture (Fig. 1B), was able to achieve a 95.7% agreement with the embryologists' consensus. In a similar spirit, Yang and colleagues propose a supervised DL model to assess microscopy image focus quality, providing an absolute quantitative measure of image focus that is independent of the observer (Yang et al., 2018). Eulenberg and colleagues use a different technique to learn discerning features to categorize cell cycle stages and identify cell state trajectories from high-throughput single-cell data (Eulenberg et al., 2017). Their DL model is trained to classify raw images into a set of discrete classes corresponding to cell cycle stages, and, through the process, learns a feature space in which data are continuously organized. When visualized using the tSNE dimensionality reduction method, feature vectors describing image data that are temporally close in their cell cycle progression are observed to be also close in feature space. A similar strategy has been used in a medical context to classify blood cell health and avoid human bias (Doan et al., 2020). DL-based algorithms have also led to improved detection of biological events such as cell division in the developing mouse embryo (McDole et al., 2018) (see Box 4). Non-DL-based ML approaches do however still offer a competitive alternative to DL, for instance in the automated identification of cell identities (Hailstone et al., 2020). There, classical ML techniques trained with smaller amounts of data and requiring less computational power than DL-based ones are shown to obtain comparatively good results.

---

**Box 4. Case study: In toto imaging and reconstruction of the early mouse embryo.**

Early development is a highly dynamic process whereby there are large changes in embryo size, shape and optical properties. Capturing the movement of cells inside the embryo and tracking cell divisions to form cell lineage maps is therefore a significant challenge, both experimentally as well as computationally. McDole and colleagues detect cell divisions using a 10-layer 4D CNN that predicts whether each voxel includes a cell division (McDole et al., 2018). The DL model is able to identify twice as many cell divisions as a human annotator, thus greatly increasing accuracy in addition to providing automation (A). The model has been trained on 11 image volumes where nuclei of both non-dividing and dividing cells were annotated, as well as 2083 annotated divisions from the entire time series. In addition, an *in toto* picture of the entire early embryo as it grows over 250× in volume is achieved by coupling custom adaptive light sheet microscopy with cell tracking. Tracking to retrieve cell fate maps is performed using a Bayesian framework with Gaussian mixture models and statistical vector flow analysis (B). Scale bar 10 μm. Image from (McDole et al., 2018).

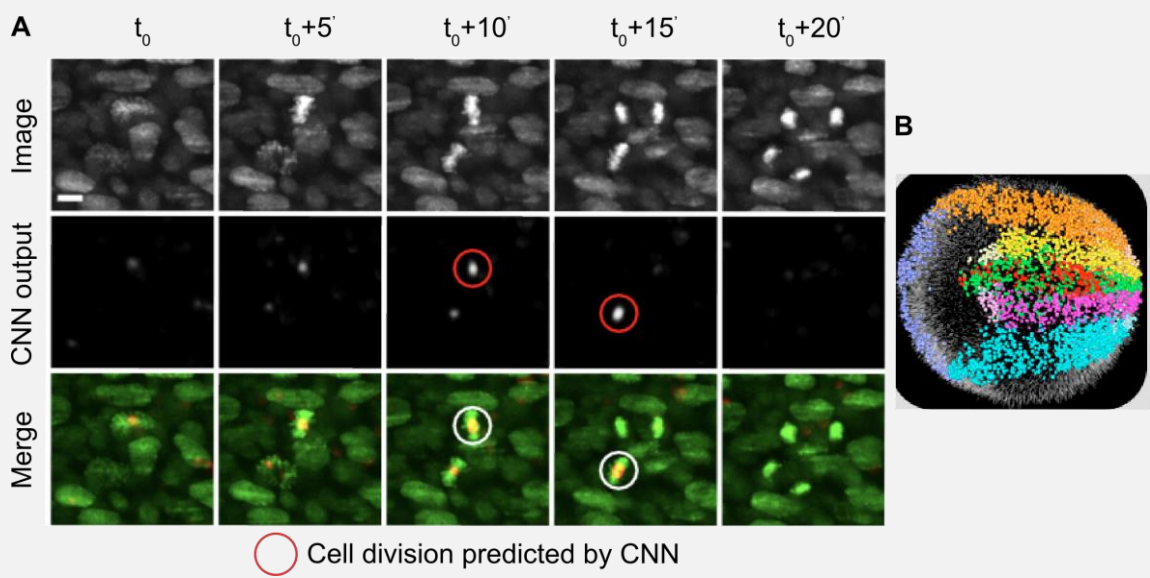

---

Although spatial tracking has been vastly studied in computer vision applications, biological objects present unique challenges. In addition to moving in and out of the field of view, cells divide, merge and can alter their appearance dramatically. Moen and colleagues propose a dedicated supervised DL approach to identify matching pairs of cells in subsequent video frames by incorporating information from surrounding frames (Moen et al., 2019). The DL model thus generates a cost matrix for all possible assignments of objects in subsequent frames, and the optimal tracking solution is retrieved with a classical combinatorial assignment algorithm, the so-called Hungarian algorithm. The full pipeline (deployed at deepcell.org) can thus automate tracking across entire populations of cells. The ability to track individual cells and follow their state, as well as that of their progeny, enables lineage reconstruction. Such a task can rapidly become manually intractable and thus greatly benefits from DL (Lugagne et al., 2020; Cao et al., 2020). One obstacle to lineage tracing is the preparation of high-quality training sets of tracks that follow cells often over long periods of time. To address this, the DL-based lineage tracing method ELEPHANT (Sugawara et al., 2021) incorporates annotation and proofreading in its user interface to reduce the need for time-consuming curated annotations. ELEPHANT can be trained on a large dataset in which only ~2% of the data are manually annotated. The model then infers on the remaining data and its predictions are validated by the user. Other methods such as 3DeeCellTracker (Wen et al., 2021) rely on simulations to build large training sets with less need for human intervention.

While cell tracking typically consists of following a single point, such as the centre of mass of a cell over time, tracking for behavioural studies requires multiple landmarks on the organism of interest. Adding landmarks to follow the movement of points on the body with respect to others is cumbersome, time consuming and not always an option. As a remedy, DeepLabCut (Mathis et al., 2018) exploits DL to automatically track points on diverse organisms from a few manual annotations. The DL model is trained on manually-annotated labels capturing striking points (e.g. left/right ear or individual digits) and learns to identify these labels on new image data without the need for added markers. DeepLabCut exploits transfer learning (see Glossary, Box 1) to achieve high accuracy tracking with small training sets of ~200 images. Other markerless tracking algorithms with a focus on several animals have been proposed to track social interactions [SLEAP (Pereira et al., 2020); id tracker ai, (Romero-Ferrero et al., 2019)] or animal posture [DeepPoseKit (Graving et al., 2019)].

## Resources and tools

The bioimage analysis community has developed a strong culture of user-friendly, open-source tool developments since its early days (Carpenter et al., 2012; Schneider et al., 2012; Eliceiri et al., 2012). Several well-established platforms, such as Weka (Arganda-Carreras et al., 2017) and ilastik (Berg et al., 2019), provide a user-friendly interface to use conventional, non-DL-based ML approaches in bioimage analysis problems. Following the rising popularity of DL in the past decade and the difficulty for non-programmers to adopt it, some of these platforms have been further developed to include DL-based algorithms, and new ones have emerged. Here, we provide an overview of selected available open-source resources developed by the bioimage analysis community that can be used to get started with DL (Table 1). Lucas and colleagues also provide an excellent in-depth discussion of open-source resources for bioimage segmentation with DL for readers wanting to explore this topic further (Lucas et al., 2021).

**Table 1.** Open-source tools for deep learning in bioimaging.

| Tool | Reference | URL | Type | Use-case | Prerequisites |
| --- | --- | --- | --- | --- | --- |
| **CellProfiler** | (McQuin et al., 2018) | cellprofiler.org | GUI-based standalone general bioimage analysis software. | Inference with pre-trained models and model training from existing ground truth for classification. | None. |
| **ilastik** | (Berg et al., 2019) | ilastik.org | GUI-based standalone general bioimage analysis software. | Inference with pre-trained models (fully supported) and model training from scratch (debug mode) for segmentation. | None. |
| **DeepImageJ** | (Gomez-de-Marisca et al., 2019) | deepimagej.github.io | ImageJ/Fiji plug-in enabling the use of pre-trained DL-based bioimage analysis algorithms. | Inference with pre-trained models for various tasks. | Experience with Fiji/ImageJ. |
| **ImJoy** | (Ouyang et al., 2019) | imjoy.io | Online computing platform for deploying DL bioimage analysis pipelines. | Inference with pre-trained models and model training from existing ground truth for various tasks. | None. |
| **ZeroCostDL4Mic** | (von Chamier et al., 2021) | github.com/HenriquesLab/ZeroCostDL4Mic/wiki | Google Colab Python notebooks implementing DL-based bioimage analysis algorithms. | Inference with pre-trained models and model training from existing ground truth for various tasks. | None. |
| **Bioimage Model Zoo** | N.A. | bioimage.io | Community-driven online repository facilitating reuse and access to pre-trained DL models. | Retrieve models architecture for various tasks, along with pre-trained weights. | Dependent on the considered model. |
| **CSBDeep** | N.A. | csbdeep.bioimagecomputing.com | Python DL toolbox for general bioimage analysis. | Model training from existing ground truth for image restoration. | Experience with Python. |

N.A., not applicable.

Several resources offer a direct point of entry into DL for bioimaging without the need for any coding expertise. The most accessible DL use-case consists of exploiting pre-trained models. This essentially means using a model that has been already trained on another image dataset to make predictions on one's own data without additional training, and requires little to no parameter tuning. Popular standalone platforms that pre-existed the DL era, such as CellProfiler ([McQuin et al., 2018](#)) and ilastik, now offer pre-trained U-net models for a variety of tasks. Both are available for all major operating systems and have their own dedicated general user interface (GUI). Since these two software packages are extremely well supported and documented, and because they contain a wealth of useful methods for image analysis in addition to DL-based ones, they probably are the lowest-entry-cost options to start experimenting with DL. Several popular pre-trained models, such as the original U-net implementation ([Ronneberger et al., 2015](#)), StarDist ([Schmidt et al., 2018](#)) and Cellpose ([Stringer et al., 2021](#)), have been made available as plug-ins for ImageJ ([Schindelin et al., 2012](#)) and napari ([napari contributors, 2019](#)). The DeepImageJ plug-in ([Gomez-de-Mariscal et al., 2019](#)), in particular, offers a unifying interface to reuse pre-trained models. A large variety of models for various image restoration and segmentation tasks are already available through it and the list is likely to grow. Searching for model implementations and pre-trained weights may be a daunting task. The reusability of most DL-based methods is significantly impacted by their custom nature, often resulting in code that is hard to distribute. The Bioimage Model Zoo ([bioimage.io](#)) is a community-driven initiative aiming to address this issue by centralizing and facilitating the reuse of published models, in DeepImageJ among others. While it is still under development and evolving quickly, the Bioimage Model Zoo is poised to evolve into a reference resource for DL models dedicated to bioimage analysis.

While pre-trained models are a good starting point, their use may not suffice to obtain good results, or worse, it may cause serious underperformance and poses a risk of generating artefacts due to dataset shift (discussed below). A more reliable, yet more involved strategy, consists of training an existing model with one's own data, either from scratch or by fine-tuning a pre-trained model, which is a particular instance of transfer learning (see Glossary, [Box 1](#)). Although several recent tools facilitate the annotation of 2 and 3D image datasets ([Hollandi et al., 2020; Borland et al., 2021](#)), the process of manually producing high-quality ground-truth annotations for training remains tedious, in particular for 3D+time datasets. The web-based platform ImJoy ([Ouyang et al., 2019](#)) hosts a large collection of plug-ins that provide interactive interfaces to generate ground-truth annotations on multi-dimensional images and train various DL algorithms. From ImJoy, algorithms can be run directly in the browser on a local host, remotely, or on a cloud server. The ZeroCostDL4Mic ([von Chamier et al., 2021](#)) toolbox also provides an excellent user-friendly solution for training DL models through guided notebooks, requiring no programming knowledge.

For the experienced programmer wishing to go further, many DL models are freely available as Python libraries. The level of user support may however dramatically vary and can range all the way from undocumented code on GitHub repositories to dedicated webpages with thorough user manuals and example data. CSBDeep ([csbdeep.bioimagecomputing.com](#)) offers one of the best examples of one such well-maintained resource, providing a wealth of documentation facilitating the reuse and adoption of DL models.

# Going further with deep learning

DL offers a plethora of exciting possibilities that go far beyond automatizing classical bioimage analysis tasks. Hereafter, we discuss some DL avenues that hold promises in the analysis of quantitative biological data beyond images and for modelling.

## Transfer learning

DL models usually require large amounts of data for training, which requires significant annotation efforts. In many cases, such ground truth sets cannot be easily generated, either because of technical limitations (e.g. in the context of image restoration) or due to the sheer amount of manual curation required. Transfer learning (see Glossary, Box 1) thus holds huge potential to enable the creation of all-rounder deep NN that can then be fine-tuned to many specific applications relying on a few annotated data only.

In the context of image restoration, (Jin et al., 2020) illustrates the benefit of transfer learning in a DL pipeline improving structured illumination microscopy image quality at low light levels. The deep NN trained with transfer learning are shown to perform equally well as their equivalents trained from scratch, but require 90% fewer ground truth samples and 10x fewer iterations to converge. Strategies aimed at reducing the number of training samples are particularly relevant to developmental biology experiments, which often rely on costly protocols to harvest few numbers of samples, resulting in particularly scarce datasets. In spite of encouraging demonstrations of the benefits of transfer learning, several questions around trust in DL-generated results remain open. Pre-trained NN must indeed be used with caution, as they may be subject to dataset shift when dealing with data that are too dissimilar to what they have been trained on. Dataset shift refers to the general problem of how information can be transferred from a variety of previous different environments to help with learning, inference, and prediction in a new one (Storkey, 2008). Understanding dataset shift thus translates to characterizing how the information held in several closely related domains (*e.g.*, data collected in other laboratories) can help with prediction in new settings. Dataset shift in bioimaging can have several origins, from batch effects to different sample preparation protocols or imaging systems. Different mitigation strategies should be used to address dataset shift depending on its nature (Quiñonero-Candela et al., 2008), and the topic is being actively investigated. However, because dataset shift is a complex phenomenon that may be hard to fully characterize, one must exercise utmost caution when using DL models outside of their training domain. When relying on pre-trained models, practitioners hold responsibility to understand the strategy and type of data that have been used to train the NN, identify the type of shift they may be facing, and remain aware of the existing mitigation strategies or lack thereof. In situations where the discrepancy between the data to be processed and those used for the initial training of the NN cannot be clearly characterized, preference should be given to conventional ML and non-DL image processing techniques.

## Style transfer

Style transfer (see Glossary, Box 1) has been famously applied in the context of artistic illustrations, allowing to turn any photographs into van Gogh paintings (Gatys et al., 2016). Similarly, it can be used to learn the image style of different microscopy modalities, with numerous applications from synthetic data generation to image enhancement. For example, this strategy has been successfully employed to adapt a nucleus segmentation task to unseen microscopy image types (Hollandi et al., 2020). In this example, style transfer is used to synthesize different types of artificial microscopy images from a single training set of ground truth labeled images. The style learned from unlabelled image samples, which are drawn from a different distribution than the training samples, is transferred to the labeled training samples. Thus, for the same set of labels, new images with realistic-looking texture, coloration and background pattern elements can be generated. This approach outperforms fine-tuning the network with a small set of additional labeled data and, in contrast, does not require any extra labelling effort. The approach has been shown to perform well on various types of microscopy images, including haematoxylin and eosin histological staining and fluorescence. Although this work focuses on nucleus segmentation, the possibility to augment difficult-to-obtain data with style transfer has enormous potential in many bioimage analysis applications beyond this specific problem.

While a lot of the enthusiasm for style transfer can be attributed to its potential as a data augmentation technique (see Glossary, Box 1), it is equally stimulating to envision it as a computational alternative *to* image acquisition. Style transfer can be exploited for synthetic image generation, for instance to produce microscopy images from different modalities, such as inferring phase-contrast microscopy images from differential interference contrast images and *vice versa* (Han & Yin, 2017). This kind of approach is appealing for many reasons, from reducing the equipment needed to reducing image acquisition time. However, current style transfer methods are oblivious to the physical properties of the specimens being imaged. Despite the visually realistic and convincing images generated for relatively simple specimens, further investigations are needed to assess how such a method would perform for more complex ones. One study perfectly formulates the underlying dilemma as 'the more we rely on DL the less confident we can be' (Hoffman et al., 2021). When doing style transfer, one specifically trains networks to lie plausibly. As such, the resulting DNN will be able to turn any object into a realistic-looking biological structure and will do so, regardless of the input; happily turning cat pictures into plausible microscopy images of cells, for example. Subsequent work is required to define confidence and uncertainty metrics for style transfer to be used in the context of scientific discovery. When exploiting style transfer strategies, biologists should make sure to follow recent developments in these directions and, most importantly, remain fully aware that a consensus of best practices on this matter has yet to be reached.

Natural texture generation

Related to style transfer, an emerging and less studied research direction with broad implications in bioimage analysis is natural texture synthesis. Texture synthesis is a well-studied problem in computer graphics where, broadly, one wants to algorithmically generate a larger image from smaller parts by exploiting geometric, regularly occurring motifs (Niklasson et al., 2021). Whether we are talking about histological images (Ash et al., 2021), early developmental patterning (McDole et al., 2019), or man-made textures such as textiles, local interactions between smaller parts (cells, morphogens or threads, respectively) can give rise to larger, emerging structures. Rather than encoding the large dimensional spaces of pixels and colours, recent DL approaches aim at describing these images in a generative way, through feed forward stochastic processes (Reinke et al., 2020; Pathak et al., 2019). The most recent approaches to generative modelling of texture synthesis are systems of partial differential equations aimed at modelling reaction–diffusion equations (Chan et al., 2020), cellular automata (Niklasson et al., 2021; Mordvintsev et al., 2020) and oscillator based, multi-agent particle systems (Ricci et al., 2021). Among these, the neural cellular automata take their inspiration from reaction–diffusion models of morphogenesis by modelling a system of locally-communicating cells that evolve and self-organise to form a desired input pattern. In the dynamic process of learning a pattern, the cells learn local rules that exhibit global properties. These rules are however abstract and not readily interpretable biologically. Mapping them to explicit gene modules involved in signaling pathways or intra cellular communication is essential to making these models useful to the developmental biological community.

Alternatively, non-black box approaches (Zhao et al., 2021) aim to identify the physical properties that can be accurately inferred from full images. Once parameters and physical properties are inferred, we can naturally wonder whether these 2D abstractions generalize, not only to 3D settings (Sudhakaran et al., 2021), but also to systems reminiscent of real cellular self-organization (Gilpin et al., 2020). In this context, an experimental counterpart is provided by studies in which the problem of pattern formation is addressed synthetically by creating morphogen systems that yield patterns reminiscent of those observed *in vivo* (Toda et al., 2020; Zhang et al., 2017).

Lessons from statistical genetics: the need for proper null models

When trying to establish whether a molecular event such as change in gene expression affects cells in an observable manner, one faces the statistical challenge of assessing significance. In past decades, statistical genetics has developed an arsenal of tools for assessing statistical significance in high dimensional problems, where hypothesis correction is essential for distinguishing between true correlation and spurious events (Barber & Candes, 2019; Stephens, 2017). Despite being focused on prediction, DL architectures for computer vision do offer many recipes for probing the interpretability of a classifier [e.g. saliency maps (Adebayo et. al, 2018)]. Additional approaches aim to identify meaningful perturbations in training data that can lead to misclassifications (Fong & Vedaldi, 2017), while others detect subparts or prototypical parts of an image that could impact classification (Chen et. al, 2019). A good resource and point of access into this vast community is the Computer Vision and Pattern Recognition Conference (CVPR) series of workshops (https://interpretablevision.github.io/index_cvpr2020.html) on interpretable ML. However, all these approaches require a large amount of training data, often unavailable in biological settings, as previously discussed.

In low data regimes, one falls back into statistician's territory and typically relies on the existence of a properly chosen null model to evaluate significance (Schäfer & Strimmer, 2005). When testing for the significance of an effect variable (e.g. gene expression change) on a quantity of interest (e.g. image feature), null models represent a way to formalize how data might look in the absence of the effect. By comparing statistical estimates from data to statistical estimates generated through appropriate null models, one can assess whether an effect is spurious or real. In simple regression models, for example, null datasets are often generated through permutation tests where, as the name suggests, data (input and outputs) are shuffled and their correlation contrasted with correlation from unshuffled data. However, permutation is not always an appropriate baseline, as illustrated in cases where data is not independently identically distributed (Dumitrascu et al., 2018; Elsayed & Cunningham, 2017). To mediate these issues, approaches for creating 'fake data' that can represent null models have been proposed with false discovery correction in mind ([Barber & Candes, 2019](#)). While fake data is easier to generate when dealing with gaussian variables, it is not clear, however, what a proper, highly structured fake image would look like. For instance, randomly shuffling the pixels of an image will just create noise. Despite the inherent difficulties, designing appropriate visual counterfactual and null models for bioimage data is essential in augmenting studies that aim to relate genomics with morphology, and is a promising area of research.

Multimodal learning

Bioimages are typically collected across multiple conditions spanning, for example, different replicates, cell types, time scales and various perturbations, such as mechanical, genetic or biochemical. Layering in additional molecular information, such as gene expression, cell lineage or chromatin accessibility ([Dries et al., 2021](#)) from high-throughput sequencing experiments, brings both the challenge of integrating data from multiple modalities, and also that of quantifying how predictive the different modalities can be of one another ([Pratapa et al., 2021](#)). Common tasks in multi-modal transfer learning particularly relevant to bioimage analysis include integrating and visualizing data from different sources (data fusion), translating between different modes (transfer), and aligning data collected across multiple modes (alignment). Data fusion is the challenge of aggregating modalities in a manner that improves prediction, especially when data might be missing or noisy. Data fusion has been thoroughly explored in the context of single-cell batch correction, where computational methods allow the integration of datasets of the same kind of modality, namely single-cell gene expression data ([Argelaguet et al., 2021](#)). However, a similar problem can be framed for microscopy data collected using different imaging modalities or in different laboratories.

Data from fundamentally different modalities, such as image data accompanied by single-cell gene expression or chromatin packing ([Clark et al., 2018, Gundersen et al., 2019](#)), poses additional challenges. Integrating them together can help in understanding whether changes in gene expression have a direct consequence not only on how individual cells look, but also on how they interact with their neighbours. However, the resolution of these different data types may be significantly different, making a direct correspondence between modalities hard to achieve ([Vergara et al., 2020](#)). In these situations, modality alignment becomes paramount ([Lopez et al., 2019](#)). A recent work integrating single-cell RNA-sequencing data and single cell nuclear imaging of naive T-cells, ([Yang et al., 2021](#)) has shown that DL representation of images contain signals predictive of true fold change of gene expression between different classes of cells.

# Conclusion

The interaction between DL and developmental biology is in its nascent stages, but will continue to grow. As a result, it is of increasing importance for biologists to become aware of applications in which DL can be exploited, but also to discern its limitations and potential pitfalls when analysing and interpreting biological data. While undeniably powerful in some settings, the use of DL comes at a cost in resources (e.g., large amount of labelled data required for training, high computational demands) and incurs risks (e.g., black-box nature of the algorithms, dataset shift). For these reasons, conventional ML and non-DL-based image processing methods should always be tried first and chosen whenever possible. As a community, cell and developmental biologists can make a conscious effort to support a scientifically sound and informed use of DL through the standardization of data acquisition, archival protocols, annotation conventions, and metadata describing image processing pipelines. There is an opportunity for the developmental biology community to centralize published data and analysis pipelines in an open source and curated manner, so as to promote a healthy use of DL in scientific discovery. As DL is rapidly pushing the limits of what is achievable in science, it calls on us to reflect on our common goals and re-evaluate how we share data and collaborate together.

# Acknowledgements

AH would like to thank J-B Lugagne for insightful discussions on DL based image analysis and acknowledges the support of Benjamin Simons and his lab. HY would like to thank Juan Caicedo for his generous DL mentorship. The authors thank Martin Weigert for providing helpful details on CARE.

# Competing interests

The authors declare no competing or financial interests.

# Funding

AH gratefully acknowledges the support of the Wellcome Trust through a Junior Interdisciplinary Research Fellowship (098357/Z/12/Z) and of the University of Cambridge through a Herchel Smith Postdoctoral Research Fellowship. He also acknowledges the support of the core funding to the Wellcome Trust/CRUK Gurdon Institute (203144/Z/16/Z and C6946/A24843). HY gratefully acknowledges that the research reported in this publication was partly supported by NIGMS of the National Institutes of Health under award number K99GM136915. BD acknowledges support from the Accelerate Programme for Scientific Discovery. VU acknowledges support from EMBL core funding.

# Supplementary Information

No supplementary information file is associated with this article.